\documentclass[12pt]{article}
\usepackage{graphicx} 

\begin{document}

\baselineskip=15pt plus 1pt minus 1pt

\begin{center}

{\large \bf  $\gamma$-soft analogue of the confined $\beta$-soft rotor model}  

\bigskip\bigskip

{Dennis Bonatsos$^{a}$\footnote{e-mail: bonat@inp.demokritos.gr},
D. Lenis$^{a}$\footnote{e-mail: lenis@inp.demokritos.gr},
N. Pietralla$^{b,c}$\footnote{e-mail: pietrall@ikp.uni-koeln.de}
P. A. Terziev$^{d}$\footnote{e-mail: terziev@inrne.bas.bg} },
\bigskip

{$^{a}$ Institute of Nuclear Physics, N.C.S.R.
``Demokritos''}

{GR-15310 Aghia Paraskevi, Attiki, Greece}

{$^{b}$ Institut f\"ur Kernphysik, Universit\"at zu K\"oln, 50937  K\"oln,
Germany} 

{$^{c}$ Institut f\"ur Kernphysik, Technische Universit\"at Darmstadt, 
64289 Darmstadt, Germany} 

{$^{d}$ Institute for Nuclear Research and Nuclear Energy, Bulgarian
Academy of Sciences }

{72 Tzarigrad Road, BG-1784 Sofia, Bulgaria}

\end{center}

\bigskip\bigskip
\centerline{\bf Abstract} \medskip 

A $\gamma$-soft analogue of the confined $\beta$-soft (CBS) rotor model 
is developed, by using a $\gamma$-independent displaced infinite well 
$\beta$-potential in the Bohr Hamiltonian, for which exact separation of
variables is possible. Level schemes interpolating between the E(5) critical 
point symmetry (with $R_{4/2}=E(4_1^+)/E(2_1^+)=2.20$)
and the O(5) $\gamma$-soft rotor (with $R_{4/2}=2.50$)
are obtained, exhibiting 
a crossover of excited $0^+$ bandheads which leads to agreement with the 
general trends of $0_2^+$ states in this region and is observed
experimentally in $^{128,130}$Xe.  

\section{Introduction} 

Critical point symmetries \cite{IacE5,IacX5}, related to shape/phase 
transitions, have attracted recently considerable attention in nuclear 
structure, since they provide parameter-free (up to overall scale factors)
predictions supported by experimental evidence 
\cite{CZE5,CZX5,ClarkE5,ClarkX5}. The E(5) critical point 
symmetry \cite{IacE5}, in particular, is related to the shape/phase transition
between vibrational [U(5)] and $\gamma$-unstable [O(6)] nuclei, while X(5) 
\cite{IacX5} is related to the transition between vibrational and 
axially symmetric prolate [SU(3)] nuclei. A systematic study of phase 
transitions in nuclear collective models has been given in 
\cite{RoweI,RoweII,RoweIII}.  

In both the E(5) and X(5) models, exact [in E(5)] or approximate
[in X(5)] separation of the $\beta$ and $\gamma$ collective variables 
of the Bohr Hamiltonian \cite{Bohr} is 
achieved, and an infinite square well potential in $\beta$ is used.
(Various analytic solutions of the Bohr Hamiltonian have been recently 
reviewed in Ref. \cite{Fortun}, while a recently introduced
\cite{RoweIV,RoweV,RoweVI} 
computationally tractable version of the Bohr collective model 
is already  in use \cite{Caprio}. )
Models interpolating between E(5) [or X(5)] and U(5) have been obtained 
by using $\beta^{2n}$ potentials (with $n=1$, 2, 3, 4) in the 
relevant [E(5) or X(5)] framework \cite{Ariasb4,E5,X5}, while an 
interpolation between X(5) and the rigid rotor limit has been achieved 
in the framework of the confined $\beta$-soft (CBS) rotor model \cite{PG}, 
by using in the X(5) framework infinite square well potentials in $\beta$ 
with boundaries $\beta_M > \beta_m \geq 0$, with the case of $\beta_m=0$ 
corresponding to the original X(5) model.  
The CBS rotor model showed considerable success in describing 
transitional and strongly 
deformed nuclei in the rare earths and actinides \cite{DP,Dusling}. 

In the present work an interpolation between E(5) and the $\gamma$-soft 
rotor [O(5)] limit is achieved, by using in the E(5) framework 
$\gamma$-independent infinite square
well potentials in $\beta$ with boundaries $\beta_M > \beta_m \geq 0$.
The model contains one free parameter, $r_\beta=\beta_m/\beta_M$,  
the case with $r_\beta=0$ corresponding to the original E(5) model, 
while $r_\beta \to 1$ leads to the $\gamma$-soft rotor [O(5)] limit.  
A special case with the two lowest excited $0^+$ states being degenerate 
occurs for $r_{\beta_c} = 0.171$. Experimental examples on the E(5) side 
and on the O(5) side of $r_{\beta_c}$ are found to correspond to $^{130}$Xe
and $^{128}$Xe, respectively. The crossover of $0^+$ bandheads observed 
at $r_{\beta_c}$ is important in reproducing the experimental trends of $0_2^+$
bandheads in the $R_{4/2}=E(4_1^+)/E(2_1^+)$ region between 2.20 [E(5)]
and 2.50 [O(5)]. 

In Section 2 the calculation of the energy spectra and $B(E2)$ transition 
rates is decribed, while results are shown and compared to 
experiment in Section 3. An overall discussion of the present results  
in given in Section 4. 

\section{The model} 

We consider the Bohr Hamiltonian \cite{Bohr} 
\begin{equation}
H = -\frac{\hbar^2}{2B}\Bigl[\Delta_R + \frac{1}{\beta^2}\Delta_\Omega\Bigr] 
+ U(\beta),
\label{bohr}
\end{equation}
with
\begin{eqnarray}
\Delta_R &=& \frac{1}{\beta^4}\frac{\partial}{\partial\beta}\beta^4
\frac{\partial}{\partial\beta} =
\frac{\partial^2}{\partial\beta^2} + \frac{4}{\beta}
\frac{\partial}{\partial\beta}, \\
\Delta_\Omega &=&
\frac{1}{\sin3\gamma}\frac{\partial}{\partial\gamma}\sin3\gamma
\frac{\partial}{\partial\gamma}
- \sum_{k=1}^{3}\frac{L_k^{\prime 2}(\theta_i)}{4\sin^2(\gamma-
\frac{2\pi}{3}k)}, 
\end{eqnarray}
where $\beta$ and $\gamma$ are the usual collective coordinates, 
$L_k^\prime$ ($k=1,2,3$) are the components of angular momentum in the 
intrinsic frame, $\theta_i$ ($i=1,2,3$) are the Euler angles, 
and $B$ is the mass parameter. The potential $U(\beta)$ depends only 
on the collective coordinate $\beta$ \cite{Wilets}. 

Using the factorized wave function 
$\Psi(\beta,\gamma,\theta_i) = F(\beta)\Phi(\gamma,\theta_i)$
\cite{IacE5,Wilets} 
the Schr\"{o}dinger equation corresponding to the Hamiltonian (\ref{bohr}) 
is separated into two parts:

(a) The angular part
\begin{equation}\label{angulareq}
-\Delta_\Omega\Phi(\gamma,\theta_i) = \tau(\tau+3)\Phi(\gamma,\theta_i),
\end{equation}
where $\tau$ is the seniority quantum number and $\Delta_\Omega$
is a quadratic invariant operator of the group SO(5) \cite{Wilets,Bes}. 
A detailed discussion can be found in Ref.~\cite{Bes}. 

(b) The radial part
\begin{equation}
\frac{d^2F(\beta)}{d\beta^2} + \frac{4}{\beta}\frac{dF(\beta)}{d\beta}
+ \left[\frac{2B}{\hbar^2}\Bigl(E - U(\beta)\Bigr) 
- \frac{\tau(\tau+3)}{\beta^2}\right]F(\beta)=0.
\label{radialeq}
\end{equation}

In the radial equation (\ref{radialeq}) we consider an infinite well potential
\cite{IacE5,Wilets} confined between boundaries \cite{PG} 
at $\beta_m$ and $\beta_M$ ($0<\beta_m<\beta_M$)
\begin{equation}\label{potential}
U(\beta)=\left\{\begin{array}{ll} 0 &,\quad \beta_m \leq\beta\leq\beta_M \\
\infty &, \quad 0\leq\beta < \beta_m, \quad\beta > \beta_M . \end{array}\right.
\end{equation}

Defining $k^2=2BE/\hbar^2$ and substituting $F(\beta)=\beta^{-3/2} P(\beta)$,
Eq. (\ref{radialeq}) in the interval $\beta\in[\beta_m, \beta_M]$  takes the 
form of a Bessel equation of $\nu$th order   
\begin{equation}\label{beseq}
\beta^2 P^{\prime\prime}(\beta) + \beta P^\prime(\beta) + \bigl(k^2\beta^2 
- \nu^2\bigr)P(\beta) = 0,
\end{equation}
where $\nu=\tau + 3/2$. The boundary conditions at $\beta_m$ and $\beta_M$ are
\begin{equation}\label{bcond}
P(\beta_m)=0, \qquad P(\beta_M)=0, \qquad 0<\beta_m<\beta_M.
\end{equation}
The general solution of Eq. (\ref{beseq}) is the  cylindrical function
\begin{equation}\label{eigfun}
P(\beta) = a J_{\nu}(k\beta) + b Y_{\nu}(k\beta),
\end{equation}
where $J_{\nu}(z)$ and $Y_{\nu}(z)$ are the Bessel functions of the first 
and second kind respectively
of order $\nu=\tau + 3/2$, and ($a, b$) are constants to be determined.
The boundary conditions (\ref{bcond}) lead to a  homogenous system for ($a,b$)
\begin{eqnarray*}
a J_{\nu}(k\beta_M) + b Y_{\nu}(k\beta_M) = 0, \\
a J_{\nu}(k\beta_m) + b Y_{\nu}(k\beta_m) = 0,
\end{eqnarray*}
which has nontrivial solutions in ($a,b$) if its determinant is set to vanish
\begin{equation}\label{kzeq}
J_{\nu}(k\beta_M)Y_{\nu}(k\beta_m) - J_{\nu}(k\beta_m)Y_{\nu}(k\beta_M) = 0.
\end{equation}
In this way the boundary conditions (\ref{bcond}) lead to a discrete 
spectrum of the parameter $k$, the values of which
are the positive roots of  Eq. (\ref{kzeq}).
Eq. (\ref{kzeq}) can be written in the form \cite{PG} 
\begin{equation}\label{xzeq}
J_{\nu}(x)Y_{\nu}(r_\beta x) - J_{\nu}(r_\beta x)Y_{\nu}(x) = 0,
\end{equation}
where $x = k\beta_M$ and the parameter $r_\beta$ denotes the ratio 
$r_\beta=\beta_m/\beta_M$.
Here we consider the case in which the parameter $\beta_M$ is fixed and 
$\beta_m$ varies in the range $0<\beta_m<\beta_M$,
the ratio $r_\beta$ taking values in the interval $0<r_\beta<1$.

Let $x_{\xi\tau}^{(r_\beta)}$ be the $\xi$th positive root of 
Eq. (\ref{xzeq}), and respectively 
$k_{\xi\tau}^{(r_\beta)} = x_{\xi\tau}^{(r_\beta)}/\beta_M$ 
be the $\xi$th positive root of Eq. (\ref{kzeq}), where $\nu=\tau+3/2$.
Then the normalized eigenfunctions $P_{\xi\tau}^{(r_\beta)}(\beta)$ can be 
represented in the form
\begin{equation}
P_{\xi\tau}^{(r_\beta)}(\beta) = 
\left[A_{\xi\tau}^{(r_\beta)}\right]^{-1/2}\Bigl[
J_{\nu}(k_{\xi\tau}^{(r_\beta)}\beta) Y_{\nu}(k_{\xi\tau}^{(r_\beta)}\beta_m) -
J_{\nu}(k_{\xi\tau}^{(r_\beta)}\beta_m) Y_{\nu}(k_{\xi\tau}^{(r_\beta)}\beta)
\Bigr] ,
\label{eigfunct}
\end{equation}
where $\beta_m\leq\beta\leq\beta_M$ and $k_{\xi\tau}^{(r_\beta)}\beta_m 
= r_\beta  x_{\xi\tau}^{(r_\beta)}$.
Then the normalized solutions of Eq. (\ref{radialeq}) in the interval 
$[\beta_m, \beta_M]$ are
\begin{equation}
F_{\xi\tau}^{(r_\beta)}(\beta) = \beta^{-3/2}P_{\xi\tau}^{(r_\beta)}(\beta).
\end{equation}
The constants $A_{\xi\tau}^{(r_\beta)}$ in (\ref{eigfunct}) are obtained from 
the normalization condition
\begin{equation}\label{efnorm}
\int_{\beta_m}^{\beta_M} \beta^4 \left[F_{\xi\tau}^{(r_\beta)}(\beta)\right]^2
\,d\beta =
\int_{\beta_m}^{\beta_M} \beta \left[P_{\xi\tau}^{(r_\beta)}(\beta)\right]^2
\,d\beta = 1. 
\end{equation}
The corresponding energy spectrum is
\begin{equation}
E_{\xi\tau}(r_\beta) = \frac{\hbar^2}{2B}\,
\left[k_{\xi\tau}^{(r_\beta)}\right]^2 =
\frac{\hbar^2}{2B\beta_M^2}\,\left[x_{\xi\tau}^{(r_\beta)}\right]^2 .
\end{equation}

In the limiting case of $\beta_m\to 0$ (or $r_\beta\to 0$) the spectrum and 
eigenfunctions correspond to the E(5) critical point symmetry \cite{IacE5}.

The factorized wave functions are denoted by 
\begin{equation}
|r_\beta;\xi\tau\mu LM\rangle \equiv
\Psi_{\xi\tau\mu LM}^{(r_\beta)}(\beta,\gamma,\theta_i)
= F_{\xi\tau}^{(r_\beta)}(\beta) \Phi^{\tau\mu}_{LM}(\gamma,\theta_i),
\end{equation}
where $\tau$ is the seniority quantum number, $\mu=0,1,2,\ldots,[\tau/3]$,
and for a given value of $\mu$
the angular momentum $L$ takes values 
$L=2\rho, 2\rho-2, 2\rho-3,\ldots, \rho+1, \rho$, where $\rho=\tau-3\mu$.
The angular part of the wave function has the form \cite{Chacon}  
\begin{equation}
\Phi^{\tau\mu}_{LM}(\gamma,\theta_i) =
N_{\tau\mu L}^{-1/2}\sqrt{\frac{2L+1}{8\pi^2}}\sum_{K} 
\phi^{\tau\mu}_{LK}(\gamma) D_{MK}^{L\ast}(\theta_i),
\end{equation}
where $N_{\tau\mu L}$ is a normalization constant and
the index $K$ in the above sum takes even values in the interval $|K|\leq L$.
In the present case we consider only states which are non-degenerate 
with respect to the quantum number $L$ in the framework
of the group embedding SO(5)$\supset$SO(3).

The reduced transition probabilities $B(E2)$ for the $E2$ transitions
\begin{equation}
B(E2; \alpha_i L_i \to \alpha_f L_f) =
\frac{|\langle\alpha_f L_f\|T^{(E2)}\|\alpha_i L_i\rangle|^2}{2L_i+1},
\end{equation}
are calculated for the quadrupole operator $T^{(E2)}$ proportional 
to the collective variable $\alpha_m$
\begin{equation}
T^{(E2)}_m \propto \beta \Bigr[D^{2\ast}_{m 0}(\theta_i) \cos\gamma+
\frac{1}{\sqrt{2}}
\bigl(D^{2\ast}_{m 2}(\theta_i) + D^{2\ast}_{m -2}(\theta_i)\bigr)
\sin\gamma  \Bigr] .
\label{te2}
\end{equation}
As a result for the $E2$ transitions one has
\begin{equation}
B(E2; L_{\xi\tau\mu} \to L'_{\xi'\tau'\mu'}) 
= R^2_{\xi'\tau';\,\xi\tau}(r_\beta)\,G^2_{\tau'\mu' L';\,\tau\mu L} ,
\end{equation}
where
\begin{equation}
R_{\xi'\tau';\,\xi\tau}(r_\beta) =
\int_{\beta_m}^{\beta_M} \beta F_{\xi'\tau'}^{(r_\beta)}(\beta) 
F_{\xi\tau}^{(r_\beta)}(\beta)\,\beta^4 d\beta,
\end{equation}
and $G_{\tau'\mu' L';\,\tau\mu L}$ are geometrical factors corresponding 
to the embedding SO(5)$\supset$SO(3). 
The selection rules for the matrix elements of the quadrupole operator 
$T^{(E2)}_m$ defined in (\ref{te2}) are
$|\Delta\tau|=1$ and $|\Delta L|\leq 2$.
We stress that all wave functions, 
energy eigenvalues, and transition matrix elements are exact analytical 
solutions of the Bohr Hamiltonian for the class of potentials considered 
here. 

\section{Analytical results and comparison to experiment} 

Analytical results for energy levels and $B(E2)$ transition rates are shown 
in Fig. 1~. The main observation regards the position of the lowest excited 
$0^+$ states. In E(5) \cite{IacE5} and for low values of 
$r_\beta < r_{\beta_c} =0.171$, $0_2^+$ 
corresponds to ($\xi=2$, $\tau=0$), while $0_3^+$ is provided by ($\xi=1$, 
$\tau=3$). 
For higher values of $r_\beta > r_{\beta_c} =0.171$ the 
picture is opposite, with $0_2^+$ corresponding to ($\xi=1$, $\tau=3$), and 
$0_3^+$ given by ($\xi=2$, $\tau=0$). 
The latter eigenstate is shifted toward infinite energy as the O(5) 
limit is approached for $r_\beta \to 1$. 
The normalized $0^+$ bandheads are shown as a function of the 
$R_{4/2}=E(4_1^+)/E(2_1^+)$ ratio 
in Fig. 2. On each curve the parameter $r_\beta$ starts from 
$r_\beta=0$ on the left, gradually increasing to the right. 
The crossover of the ($\xi=2$, $\tau=0$) and the ($\xi=1$, $\tau=3$) 
curves occurs at $r_{\beta_c}=0.171$.

The existence of the $0^+$ state's crossover is crucial in keeping the 
model predictions
in agreement with the general trends shown by the experimental 
$R_{0/2}=E(0_2^+)/E(2_1^+)$ ratio as a function of the $R_{4/2}
=E(4_1^+)/E(2_1^+)$ ratio, given in Ref. \cite{Chou}. In the region 
with $2.20<R_{4/2}<2.50$, covered by the present model, the experimental 
$R_{0/2}$ values stay indeed below 5.0, in agreement with what is seen in 
Fig.~2 
for the ($\xi=1$, $\tau=3$) bandhead. 

It is interesting to identify nuclei corresponding to parameter values 
near the region $r_\beta=0.15$-0.20, in which the crossover of the 
$0^+$ bandheads occurs. Below $r_{\beta_c}=0.171$ the situation resembles 
the one in E(5), with $6_1^+$, $4_2^+$ and $0_3^+$ states being nearly 
degenerate, 
while the $0_2^+$ state lies lower than them. 
Beyond $r_{\beta_c}=0.171$ the near degeneracy applies to the 
$6_1^+$, $4_2^+$ and $0_2^+$ states, while the $0_3^+$ state lies 
higher than them. 
This situation occurs in the neighboring nuclei 
$^{130}$Xe (corresponding to $r_\beta=0.12$) and $^{128}$Xe (reproduced by 
$r_\beta=0.21$), shown in Fig. 3. (The parameter $r_\beta$ has been fitted 
to the experimental $R_{4/2}$ ratio of each nucleus.) In the latter case, 
known $B(E2)$ values, too, agree remarkably well with the theoretical 
predictions. 

Both before and after the crossover, the $0^+$
bandhead with ($\xi=1$, $\tau=3$) is connected by a strong $E2$ transition
to the $2_2^+$ state, while the $0^+$ bandhead with ($\xi=2$, $\tau=0$) 
decays less strongly to the $2_1^+$ level. These interband $B(E2)$ 
values provide a stringent test to the model. 
However, their absolute values are unknown experimentally. 

It is nevertheless possible to unambiguously characterize the
predominant nature of the two excited $0^+_{2,3}$ states of
$^{128,130}$Xe by considering their $E2$ decay branching
ratios to the lowest two $2^+_{1,2}$ states.
We thus define the double-ratio
\begin{eqnarray}
{\cal Z}(0^+_{3/2}) & = &
\frac{B(E2; 0^+_3 \to 2^+_2)/B(E2; 0^+_3 \to 2^+_1)}
     {B(E2; 0^+_2 \to 2^+_2)/B(E2; 0^+_2 \to 2^+_1)} \\
  & = &
\frac{
 \left(
  \frac{
   E_\gamma(0^+_3 \to 2^+_1)}
   {E_\gamma(0^+_3 \to 2^+_2)
  }
 \right)^5
 \frac{I_\gamma(0^+_3\to 2^+_2)}{I_\gamma(0^+_3\to 2^+_1)}}
{ \left(
  \frac{
   E_\gamma(0^+_2 \to 2^+_1)}
   {E_\gamma(0^+_2 \to 2^+_2)
  }
 \right)^5
 \frac{I_\gamma(0^+_2\to 2^+_2)}{I_\gamma(0^+_2\to 2^+_1)}}
\end{eqnarray}
for which one expects values $> 1$ for $r_\beta < r_{\beta_c}$
and $< 1$ for $r_\beta > r_{\beta_c}$, respectively.
Eq. (23) involves $\gamma$-ray energies and intensity ratios. 
The data \cite{130Xe,128Xe,Miyahara} yield values of
${\cal Z}(0^+_{3/2}) = 52 \pm 30$ for $^{130}$Xe and
${\cal Z}(0^+_{3/2}) = 0.32 \pm 0.17$ for $^{128}$Xe, 
as given on Table 1. The experimental values for $^{130}$Xe and $^{128}$Xe 
differ by two orders of magnitude. 
Despite the large uncertainties that originate in the 50\% 
uncertainty for the low intensity of the initially forbidden
$0^+ \to 2^+_2$ low-energy transition \cite{130Xe,128Xe,Miyahara}, 
the data prove the crossing of the different $0^+$
configurations with ($\xi=2$, $\tau=0$) and ($\xi=1$, $\tau=3$)
between $^{130}$Xe and $^{128}$Xe, as predicted by the model 
from a fit to the relative $4^+_1$ excitation energy $R_{4/2}$. 

Having identified the $0^+$ configuration crossing we can 
analyze it quantitatively in a two-state mixing 
scenario. 
These close-lying experimental $0^+$ states are considered as an 
orthogonal mixture of the crossing model states
$0^+_2 = \alpha_2 0^+_{\xi=2,\tau=0} + \alpha_3 0^+_{\xi=1,\tau=3}$ and 
$0^+_3 = -\alpha_3 0^+_{2,0} + \alpha_2 0^+_{1,3}$ due 
to residual interactions not accounted for by the simple model. 
This yields 
${\cal Z}(0^+_{3/2}) = (\alpha_2/\alpha_3)^4$ due to the $\tau$-selection 
rules for the $E2$ operator in the unperturbed situation 
and allows for the mixing coefficients to be determined. 
The squared amplitudes $\alpha_2^2$ quantify the $0^+$ configuration 
crossing. 
The contribution of the $0^+_{\xi=2,\tau=0}$ model state to the 
observed $0^+_2$ state drops from 88(3)\% in $^{130}$Xe 
to 43(3)\% in $^{128}$Xe as displayed in Table \ref{table}. 

With that information a further test of the model prediction for 
$E2$ transition rates can be performed assuming only the validity 
of the two-state-mixing scenario. 
The relative strengths of the unperturbed interband $E2$ transitions 
can be extracted from the experimental $E2$ branching ratios 
\begin{eqnarray}
\label{eq:E2test} 
& & \left[\frac{B(E2; 0^+_{1,3} \to 2^+_2)}
           {B(E2; 0^+_{2,0} \to 2^+_1)}\right]_{\rm unperturbed} \\ 
\nonumber 
& = & 
  \left(\frac{\alpha_3}{\alpha_2}\right)^2 
       \left[\frac{B(E2; 0^+_3 \to 2^+_2)}
                  {B(E2; 0^+_3 \to 2^+_1)}\right]_{\rm expt} \\ 
\nonumber 
& = & \sqrt{
        \frac{
          \frac{I_\gamma(0^+_2\to 2^+_2)}{I_\gamma(0^+_2\to 2^+_1)} 
          \left(\frac{E_\gamma(0^+_2 \to 2^+_1)}
                     {E_\gamma(0^+_2 \to 2^+_2)}\right)^5
             }{
          \frac{I_\gamma(0^+_3\to 2^+_1)}{I_\gamma(0^+_3\to 2^+_2)} 
          \left(\frac{E_\gamma(0^+_3 \to 2^+_2)}
                     {E_\gamma(0^+_3 \to 2^+_1)}\right)^5
             }
           }
\end{eqnarray} 
The rhs involves $E2$ intensity ratios in the perturbed (experimental) 
situation. 
The experimental values (rhs) are compared to the theoretical 
values for the lhs of Eq.~(\ref{eq:E2test}) 
at the bottom of Table \ref{table}. 
The data 
on $^{128}$Xe 
coincide with the model within the uncertainties. 
The data on $^{130}$Xe also exhibit the predicted dominance of 
the $B(E2; 0^+_{1,3} \to 2^+_2)$ value over the 
$B(E2; 0^+_{2,0} \to 2^+_2)$ value but by an order of magnitude 
more pronounced than theoretically expected. 
This numerical deviation calls for better data on the weak 
$0^+_2 \to 2^+_2$ 672-keV decay intensity with its present 
uncertainty of 50\% \cite{Miyahara}. 

\section{Summary} 

A $\gamma$-soft analogue of the confined $\beta$-soft (CBS) rotor model 
has been constructed and exactly solved analytically, 
by using a $\gamma$-independent displaced infinite well 
$\beta$-potential in the Bohr equation, in which exact separation of variables
is possible in this case. The model obtained contains one free parameter,
the ratio $r_\beta=\beta_m/\beta_M$ of the positions of the left wall 
($\beta_m$) and the right wall ($\beta_M$) of the potential well, and 
interpolates between the
E(5) critical point symmetry, possessing $R_{4/2}=E(4_1^+)/E(2_1^+)=2.20$
and obtained for $r_\beta=0$, 
and the $\gamma$-soft rotor O(5), having $R_{4/2}=2.50$ and obtained 
for $r_\beta \to 1$. 
Due to the explicit O(5) symmetry the model might be addressed as the 
{\sl O(5)-Confined $\beta$-Soft Rotor Model} [O(5)-CBS]. 
A crossover of excited $0^+$ bandheads as a function of $R_{4/2}$
is predicted, which is crucial in keeping 
the model predictions for the $0_2^+$ bandhead in good agreement with 
experimental systematics in this region of $R_{4/2}$ ratios. This crossover
is manifested in $^{128,130}$Xe as is seen quantitatively
from experimental $E2$ decay intensity ratios. 
Information on relative and absolute 
$E2$ transitions in $^{128}$Xe are in good agreement with the model 
predictions when simple configuration mixing is accounted for, 
while more accurate experimental information 
on $B(E2)$s in $^{130}$Xe is desirable for further significant tests 
of the model.

\begin{table}[b]
\small
\caption{Comparison of data \cite{130Xe,128Xe,Miyahara} on $^{130,128}$Xe 
         to the model [O(5)-CBS] and the two-state mixing scenario (see 
Section 3).}
\label{table}
\hrule
\begin{tabular}{c|cc|cc}
  & $^{130}$Xe & O(5)-CBS & $^{128}$Xe & O(5)-CBS \\[1mm] 
  &       & $r_\beta = 0.12$ &    & $r_\beta = 0.21$ \\[1mm] 
${\cal Z}(0^+_{3/2})$ & 52(30) & $\infty$ & 0.32(17) & 0 \\[1mm] 
$\alpha_2^2$ & 0.88(3) & -- & 0.43(3) & -- \\[1mm]
$\frac{B(E2; 0^+_{1,3} \to 2^+_2)}{B(E2; 0^+_{2,0} \to 2^+_1)}$ & 
 28(9) & 2.8 & 3.7(1.0) & 3.8 
\end{tabular}
\hrule 
\end{table}

\begin{figure*}
\includegraphics[width=75mm]{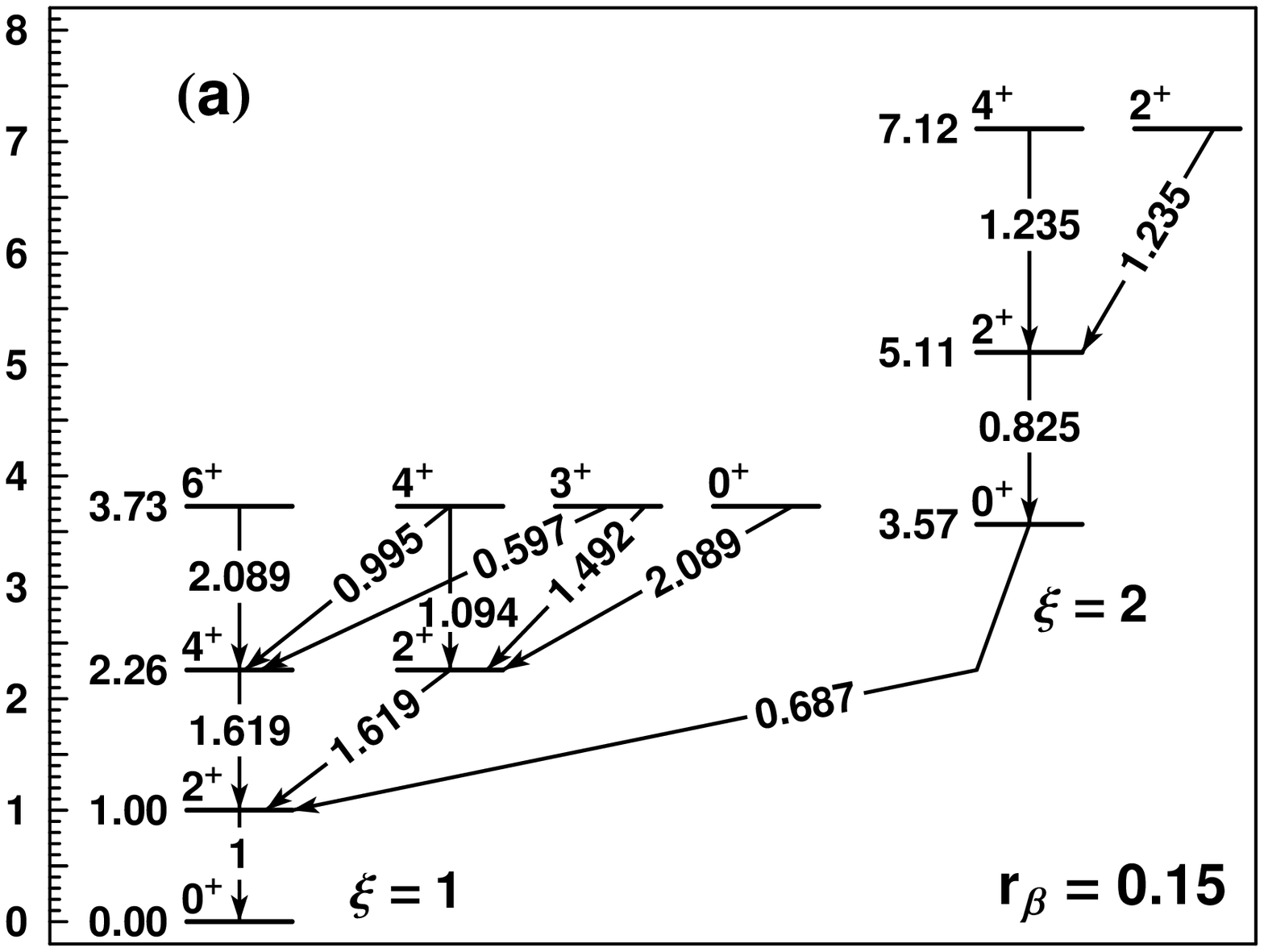}\hspace{10mm}
\includegraphics[width=75mm]{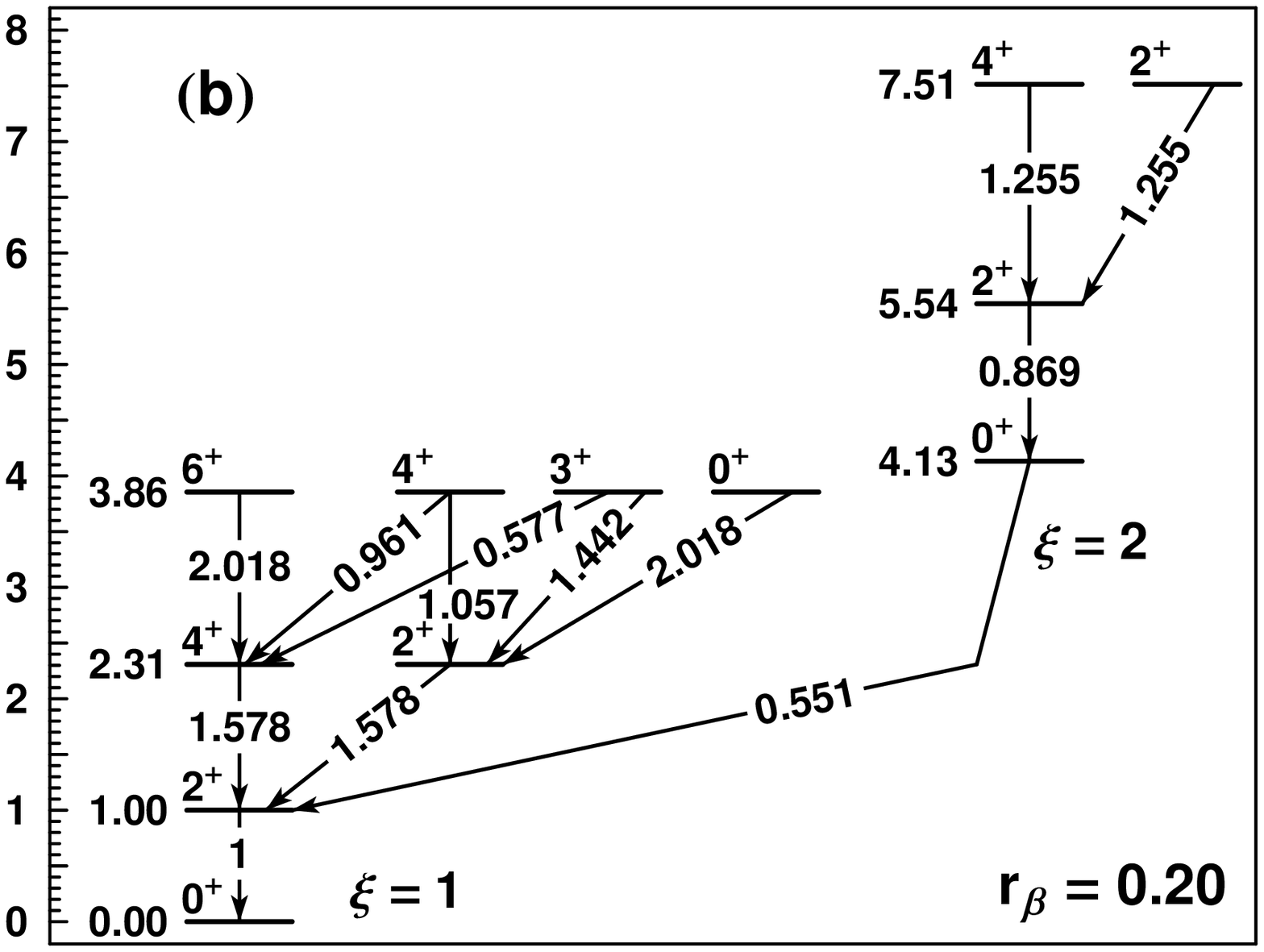}
\caption{Energy levels (normalized to the excitation energy of the $2_1^+$
state)
and $B(E2)$ values (normalized to $B(E2; 2_1^+\to 0_1^+)$)
for two different values of the structural parameter $r_\beta=\beta_m/\beta_M$
around the $0^+_{2,3}$ crossing.}
\end{figure*}

\bigskip

\begin{figure}
\includegraphics[width=75mm]{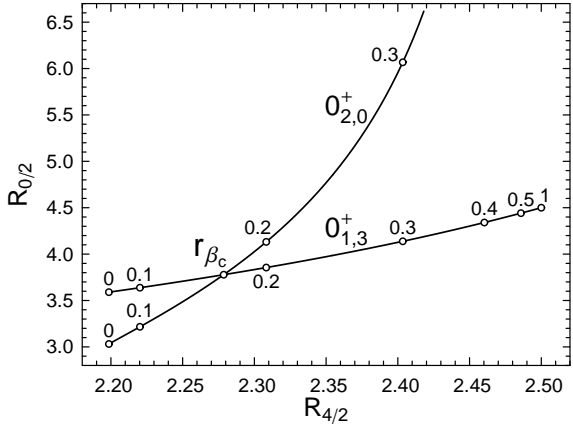}
\caption{Energy of $0^+$ states (normalized to the energy of the $2_1^+$
state), labeled as $R_{0/2}$, vs. the ratio $R_{4/2}=E(4_1^+)/E(2_1^+)$.
The parameter $r_\beta$ on each curve starts from zero at the left end,
increasing to the right.
The crossover of the ($\xi=2$, $\tau=0$, $L=0$) and the
($\xi=1$, $\tau=3$, $L=0$) curves occurs at
$r_{\beta_c} \approx 0.171263$,
$(R_{4/2},\ R_{0/2})_{c} \approx (2.27861,\ 3.77797)$.}
\end{figure}

\bigskip

\begin{figure*}
\includegraphics[width=75mm]{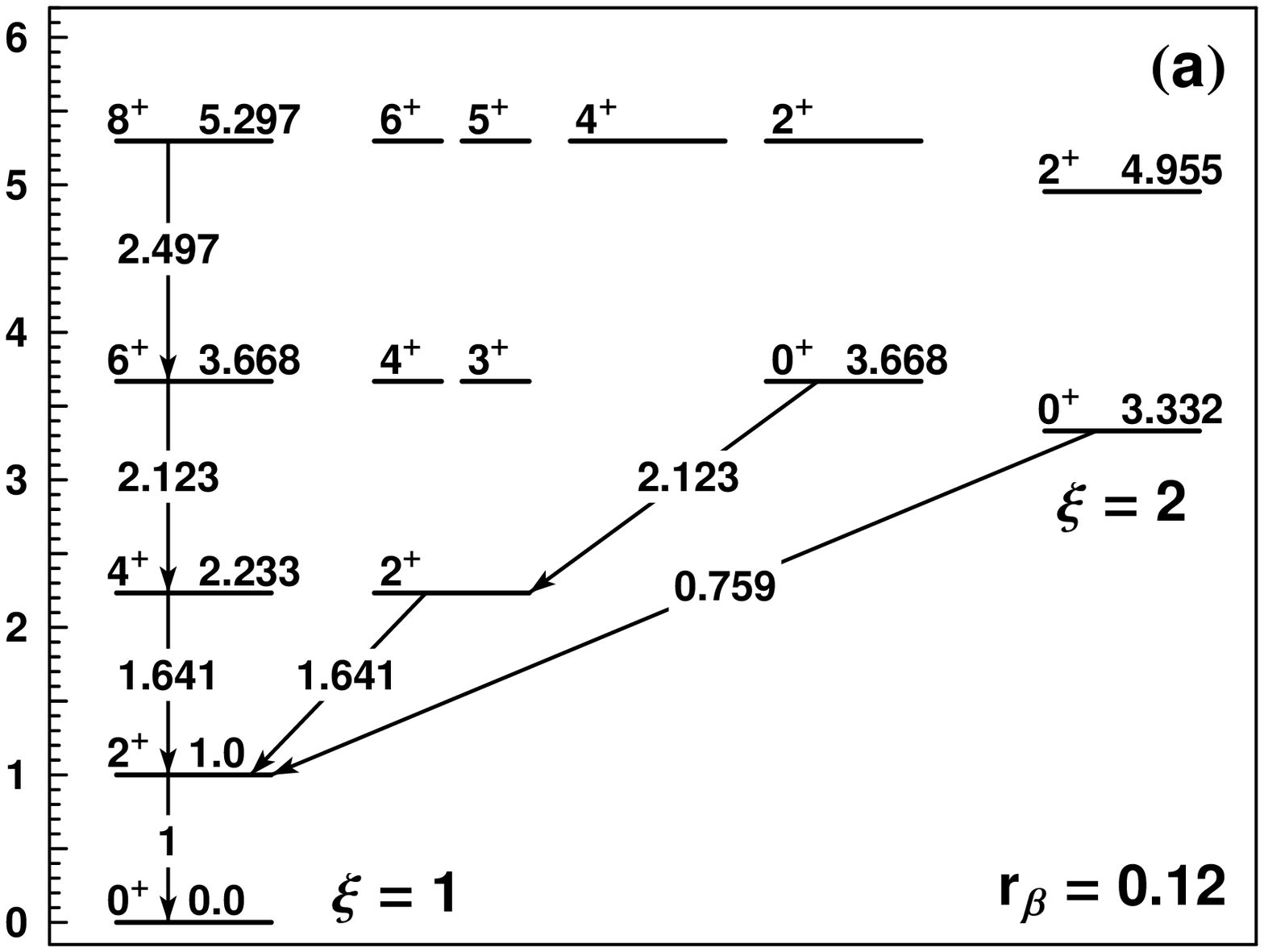}\hspace{10mm}
\includegraphics[width=75mm]{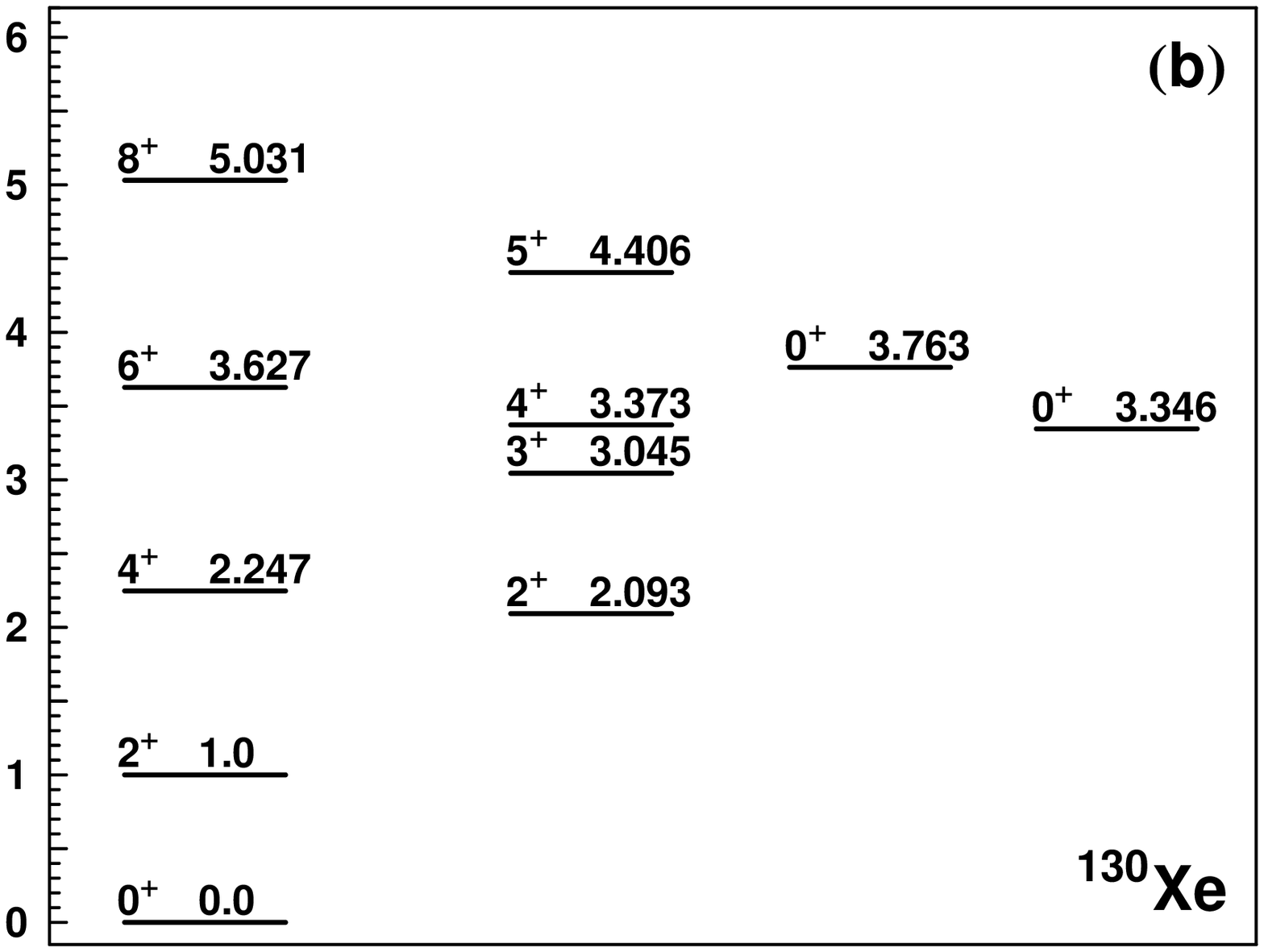}\\[1mm]
\includegraphics[width=75mm]{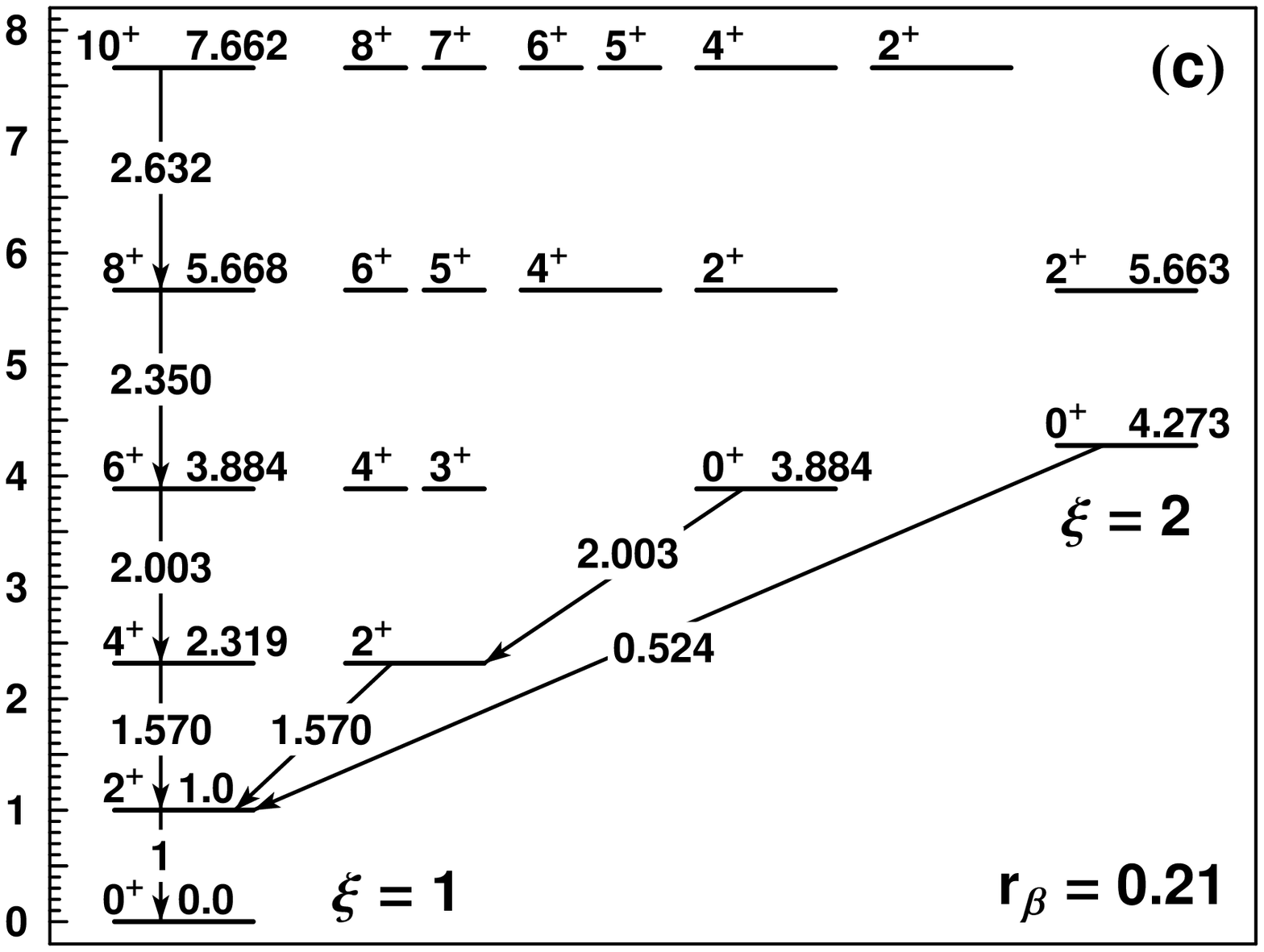}\hspace{10mm}
\includegraphics[width=75mm]{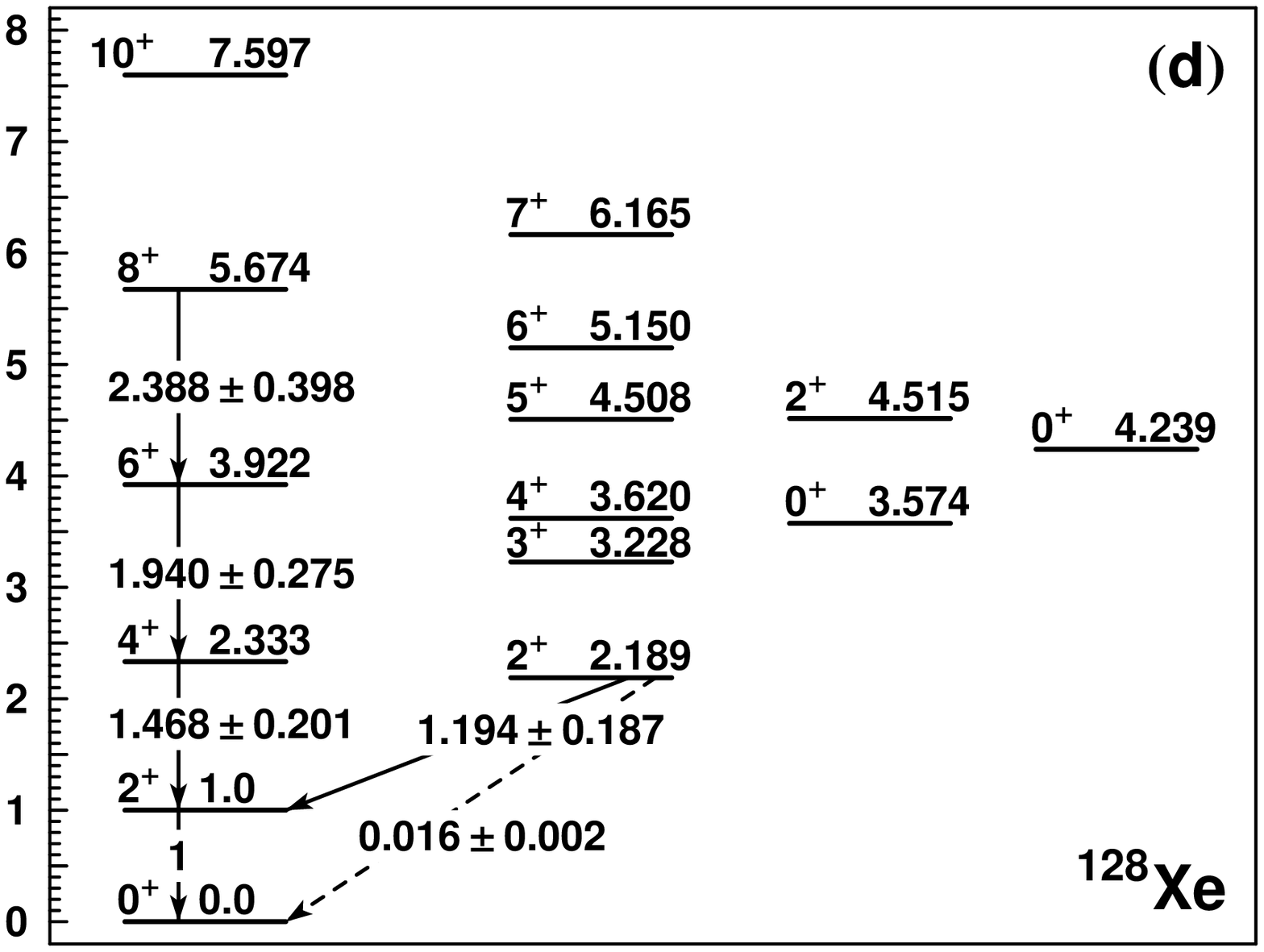}
\caption{Comparison of the model predictions for $r_\beta=0.12$ (a) to the
experimental
data of $^{130}$Xe [26] (b), and of the model predictions for
$r_\beta=0.21$ (c) to the experimental data for $^{128}$Xe [27] (d).}
\end{figure*}
\end{document}